\documentclass[aps,prl,twocolumn,showpacs,groupedaddress]{revtex4}

\usepackage[centertags]{amsmath}
\usepackage{graphicx}
\usepackage{amsfonts}
\usepackage{epstopdf}
\usepackage{ulem}
\usepackage{color}

\begin{document}

\title{Hydrodynamic Supercontinuum}

\author{A. Chabchoub$^{1,2\ast}$, N. Hoffmann$^{2,3}$, M. Onorato$^{4,5}$, G. Genty$^6$, J. M. Dudley$^7$, and N. Akhmediev$^8$}
\affiliation{$^1$ Centre for Ocean Engineering Science and Technology, Swinburne University of Technology, Hawthorn, Victoria 3122, Australia}
\affiliation{$^2$ Department of Mechanical Engineering, Imperial College London, London SW7 2AZ, United Kingdom}
\affiliation{$^3$ Dynamics Group, Hamburg University of Technology, 21073 Hamburg, Germany}
\affiliation{$^4$ Dipartimento di Fisica, Universit\`a degli Studi di Torino, 10125 Torino, Italy}
\affiliation{$^5$ Istituto Nazionale di Fisica Nucleare, INFN, Sezione di Torino,  10125 Torino, Italy}
\affiliation{$^6$ Tampere University of Technology, Optics Laboratory, 33101 Tampere, Finland}
\affiliation{$^7$ Institut FEMTO-ST, UMR 6174 CNRS- Universit\'{e} de Franche-Comt\'{e}, 25030 Besan\c{c}on, France}
\affiliation{$^8$ Optical Sciences Group, Research School of Physics and Engineering, The Australian National University, Canberra ACT 0200, Australia}
\email{achabchoub@swin.edu.au}
\begin{abstract}
We report the experimental observation of multi-bound-soliton solutions of the nonlinear Schr\"odinger equation (NLS) in the context of hydrodynamic surface gravity waves. Higher-order $N$-soliton solutions with $N=2,3$ are studied in detail and shown to be associated with self-focussing in the wave group dynamics and the generation of a steep localized carrier wave underneath the group envelope. We also show that for larger input soliton numbers the wave group experiences irreversible spectral broadening, which we refer to as a hydrodynamic supercontinuum by analogy with optics. This process is shown to be associated with the fission of the initial multi-soliton into individual fundamental solitons due to higher-order nonlinear perturbations to the NLS. Numerical simulations using an extended NLS model described by the modified nonlinear Schr\"odinger equation (MNLS), show excellent agreement with experiment and highlight the universal role that higher-order nonlinear perturbations to the NLS play in supercontinuum generation.
\end{abstract}
\maketitle
The generation of new frequency components is a defining feature of nonlinear physics. Indeed, perhaps the most spectacular phenomenon of nonlinear physics occurs when a narrow band input wave group undergoes rapid spectral broadening as a result of strong nonlinear interactions to create a broadband spectrum. Such spectral broadening has been particularly studied in an optical context, where the interaction between an intense electromagnetic pulse and a nonlinear medium can generate a quasi-continuous broadband spectrum known as a supercontinuum \cite{dudley2006supercontinuum,DudleyTaylor,dudleygenty}. Much insight into the physics of supercontinuum generation has been obtained using models of the underlying wave propagation described by the NLS \cite{zakharov68}, along with extensions to include higher-order dispersive and nonlinear perturbations \cite{dudley2006supercontinuum}. 
Besides optical waves, an important feature of the NLS model is that it provides a general description of a wide range of weakly nonlinear dispersive systems such as Langmuir waves in an unmagnetized plasma, Bose-Einstein condensates and deep-water surface gravity waves \cite{belashov2005solitary}. In fact, for surface gravity waves, the NLS equation was derived more than 40 years ago \cite{zakharov68} and was shown to be integrable via the Inverse Scattering Transform \cite{zakharov72}.

A variety of exact solutions have since been presented, with the most celebrated being the propagation-invariant bright soliton first observed experimentally in the late 1970's \cite{yuen1982nonlinear}. More recently, breather solutions of the NLS have attracted significant attention as it has been suggested that {\it breathers on finite background} \cite{kuznetsov1977solitons,akhmediev1985generation,peregrine1983water} can be considered as prototypes of the rogue waves \cite{OnoratoReport} observed on the surface of the ocean \cite{kharif2009}.  Such breathers on finite background have now been seen experimentally under controlled conditions in a range of systems including optics \cite{kibler2010peregrine,kibler2012observation}, plasma physics \cite{bailung2011observation}, and hydrodynamics \cite{chabchoub2011rogue,chabchoub2012super}.
In a hydrodynamical context, a different class of NLS in the form of {\it breathers on zero-background} has, however, received less attention \cite{satsuma1974b,akhmediev1997solitons,akhmediev1993spatial}. In fact, it is very surprising that such solutions generally known as higher-order solitons or Satsuma-Yajima breathers and which have been seen in optics more than 30 years ago \cite{mollenauer1980experimental} have never been observed in wave tank laboratory experiments. Higher-order solitons can be considered as the nonlinear superposition of multiple fundamental solitons with evolving relative phases such that recurrent cycles of envelope compression and expansion are observed over a characteristic distance scale known as the soliton period. Significantly, it is the initial compression stage of higher-order soliton propagation that is important in optical supercontinuum generation, as the associated large spectral broadening causes deviation from ideal NLS dynamics, and induces fission into multiple fundamental solitons \cite{dudley2006supercontinuum}. 

In this Letter, we present the experimental observations of higher-order solitons in a water wave tank, and show that, in the perturbative regime of hydrodynamic nonlinearity, a higher-order water wave soliton can also split into fundamental solitons and generate a broad and continuous spectrum in the same way as seen in optics. By analogy with the corresponding optical phenomenon, we describe this as a hydrodynamic supercontinuum. Significantly, we show that the observed wave dynamics and fission mechanism are well described within the framework of an extended NLS model, the MNLS, with the distance at which soliton fission occurs scaling in the same way as in optics.  Our results show clearly that the essential physics of water wave propagation remains well-described by this model over a wide range of experimental parameters and indeed suggest that fission of multi-soliton bound state and associated supercontinuum generation may be a universal phenomenon encountered in a wide range of nonlinear systems governed by perturbed NLS-type equations.

Our analysis is based on the focusing NLS equation appropriate for describing deep-water wave packet evolution in space $x$ \cite{osborne2010nonlinear}:
\begin{eqnarray}
i\left(\frac{\displaystyle\partial A}{\displaystyle \partial
x}+\frac{\displaystyle2k_0}{\displaystyle\omega_0}\frac{\displaystyle\partial A}{\displaystyle \partial
t}\right)-\frac{\displaystyle k_0}{\displaystyle
\omega_0^2}\frac{\displaystyle\partial^2A}{\displaystyle \partial
t^2}-k_0^3\left|A\right|^2A=0. \label{NLS}
\end{eqnarray}
Here $k_0$ represents the carrier wave number and $\omega_0$ is the corresponding
angular frequency related to $k_0$ {\it via} the dispersion relation $\omega_0=\sqrt{g k_0}$, where $g$ denotes the gravitational acceleration. Deep-water wave packets propagate with the group velocity:
$c_g=\frac{\displaystyle\operatorname{d}\omega}{\displaystyle\operatorname{d}k}\Big\vert_{k=k_0}=\frac{\displaystyle\omega_0}{\displaystyle2 k_0}.$
Taking into account the second Stokes harmonic, the surface elevation $\eta(x,t)$ can be represented in terms of the complex envelope $A(x,t)$ as:
\begin{align}
\eta(x,t)&=\frac{\displaystyle 1}{\displaystyle 2}\left(A\left(x,t\right)e^{i\vartheta}+\frac{\displaystyle 1}{\displaystyle 2}k_0 A^2\left(x,t\right)e^{2i\vartheta}+c.c.\right), \label{surfaceelevation}
\end{align}
where $c.c.$ denotes the complex conjugate and $\vartheta=\left(k_0x-\omega_0t\right)$.
After rescaling the space, time and amplitude variables in the equation (\ref{NLS}), one obtains the well-known dimensionless form of the NLS:
\begin{eqnarray}
i\psi_X+\psi_{TT}+2\left|\psi\right|^2\psi=0.
\label{NLSS}
\end{eqnarray}
The two-soliton solution, which corresponds to a breather on zero-background of Eq. (\ref{NLSS}) can be written analytically as \cite{satsuma1974b}:
\begin{align}
\psi_2\left(X,T\right)=&\frac{4\left(\textnormal{cosh}\left(3T\right)+3\textnormal{ cosh}\left(T\right)\exp\left(8iX\right)\right)}{\textnormal{ cosh}\left(4T\right)+4\textnormal{ cosh}\left(2T\right)+3\cos\left(8X\right)}
\\ \nonumber
&\times\exp\left(iX\right).
\end{align}
With $X=0$ in the wave tank defined as the position of the mechanical paddle generating the waves, the initial condition for exciting the two-soliton solution takes the simple hyperbolic-secant form with amplitude $\psi_2(X=0,T) = 2\operatorname{sech}(T)$. Note that general analytic forms for higher $N$ can be derived using variety of techniques \cite{akhmediev1997solitons}, but writing them in closed form becomes unpractical for orders larger than $N=3$ \cite{MollenauerGorden}. Nonetheless, the particular initial condition for exciting any $N$-soliton has the simple generic form $\psi_N(X=0,T)=N\operatorname{sech}(T)$.
% Figure 1
\begin{figure}[h]
\centering
 \includegraphics[width=7.6cm]{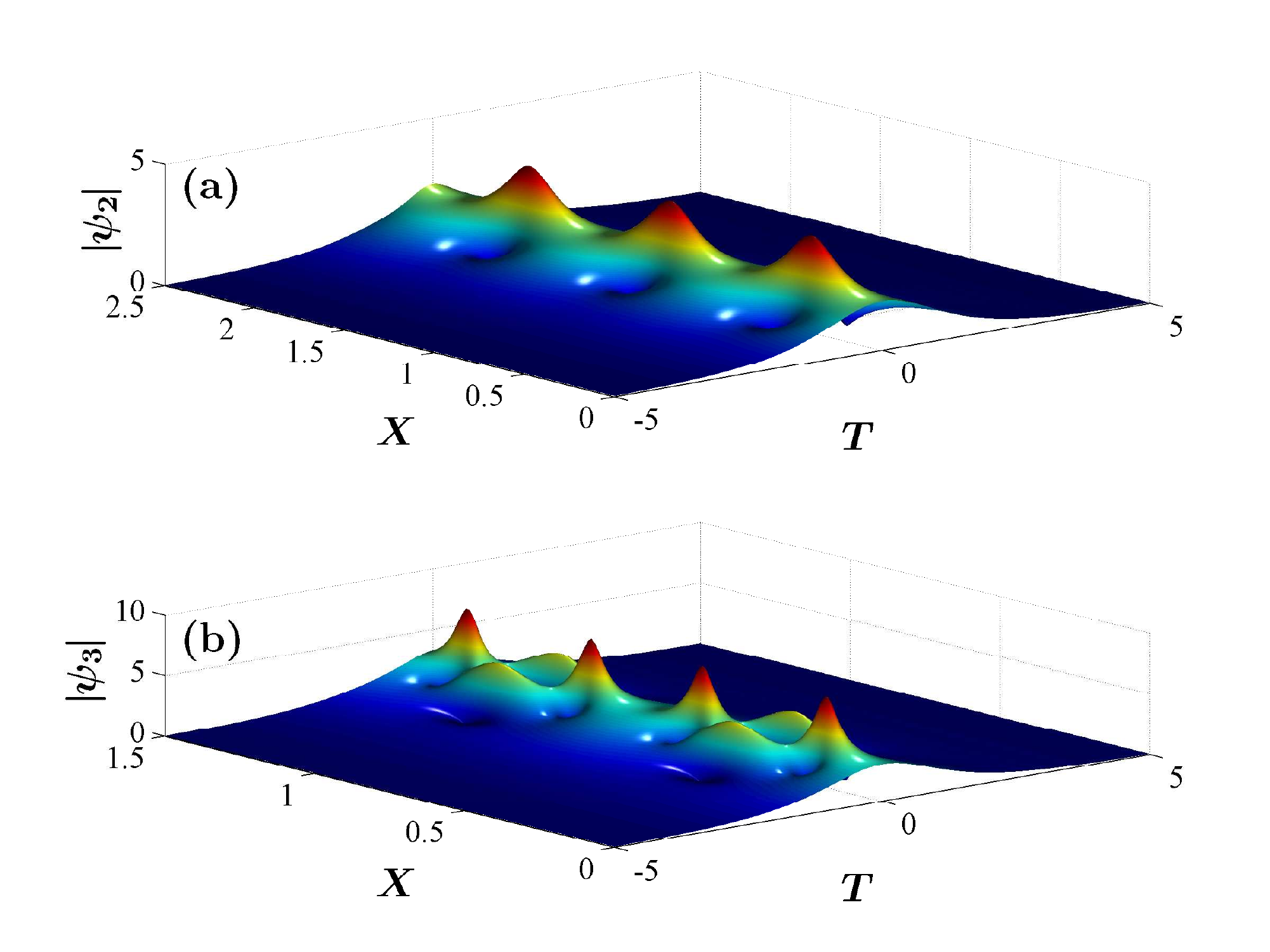}
\caption{(Color online) Theoretical evolution of the (a) $N=2$ soliton and (b)  $N=3$ soliton solution of the NLS.}\label{SYs}
\end{figure}
Figure \ref{SYs} shows the theoretical evolution of the envelope of the two- and three-soliton solutions along $X$ and $T$. Excitation of these breathers in a water tank is an important confirmation of the ability of the NLS to capture extreme localization of water waves with high amplitude features. However, we emphasize that the NLS model is only the lowest order approximation and that the dynamics of the surface waves can be influenced by higher order effects; as we discuss later, this can dramatically increase the spectral broadening during the evolution of multi-soliton solutions.
% Figure 2
\begin{figure}[h]
\centering
\includegraphics[width=5.9cm]{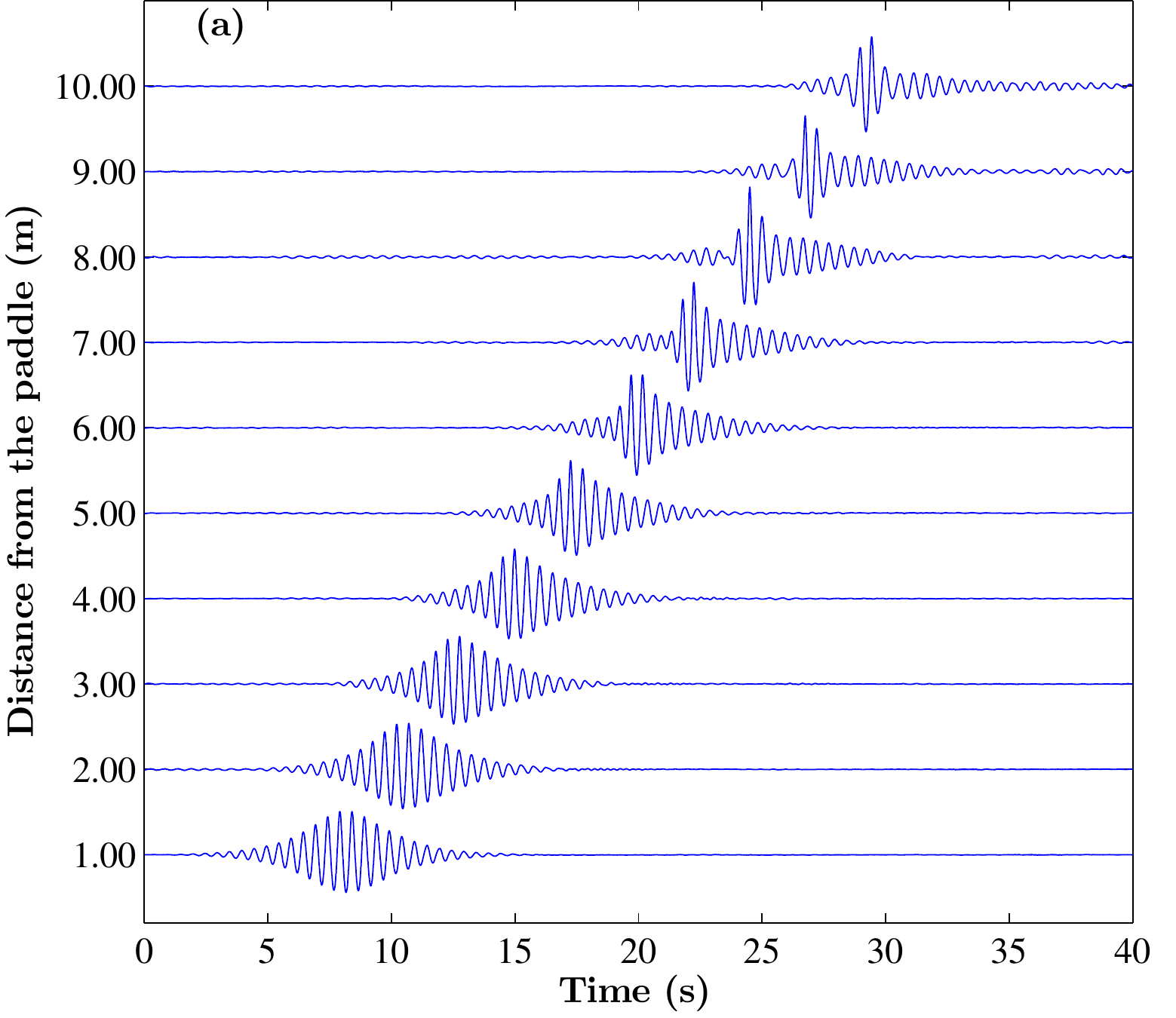}
\includegraphics[width=5.8cm]{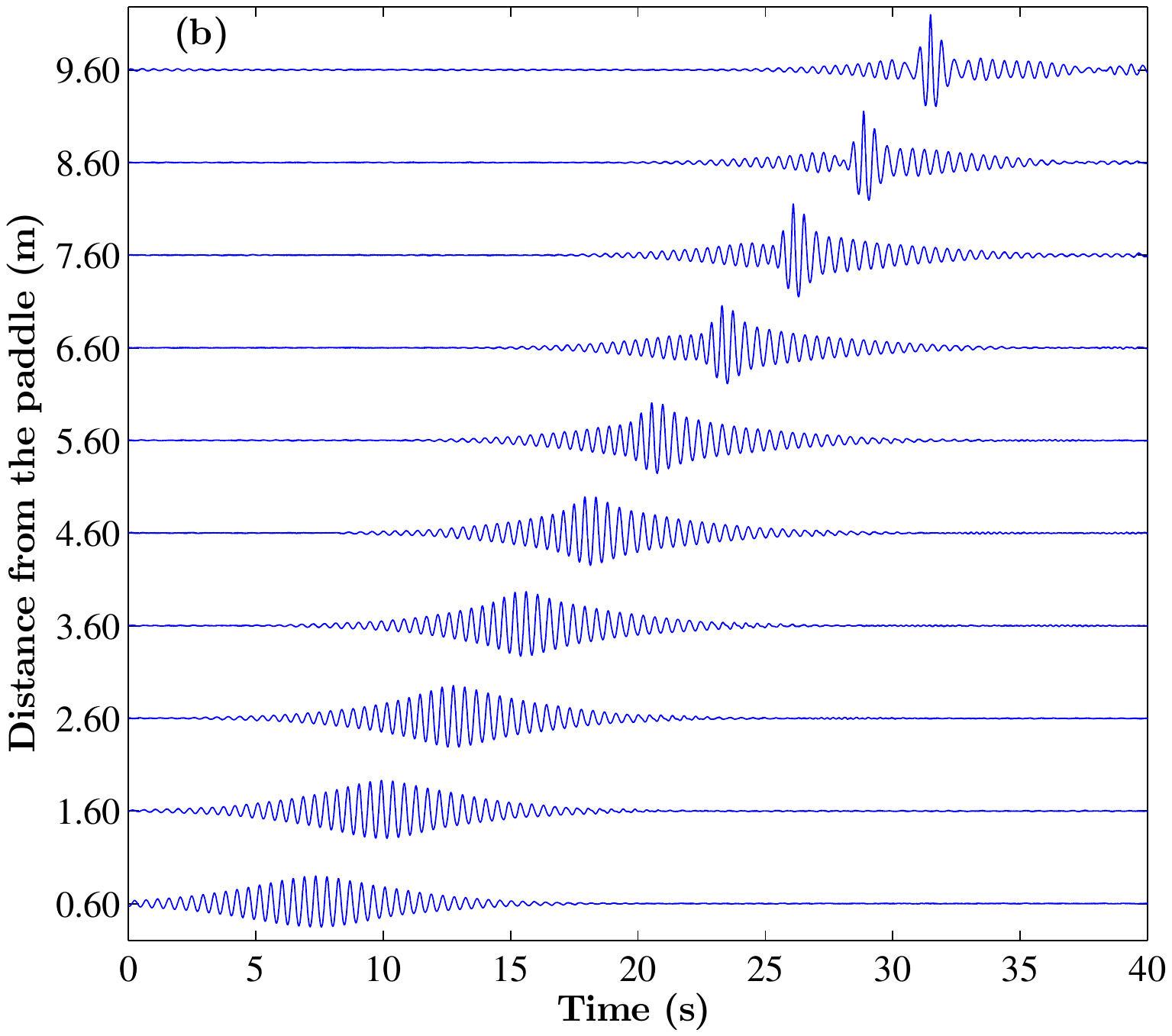}
\caption{(Color online) (a) Recorded experimental evolution of the  $N=2$ soliton along the wave flume for an initial carrier amplitude of $a_0=5$ mm and carrier steepness of $\varepsilon_0=0.08$. (b) Recorded experimental evolution of the  $N=3$ soliton along the wave flume for a carrier amplitude of $a_0=2$ mm and carrier steepness of $\varepsilon_0=0.04$. The gauges are equally spaced by 20 cm along the wave tank but for clarity we only plot the evolution at selected distances.}\label{evo}
\end{figure}

Our experimental setup is described in \cite{chabchoub2012super}. The initial condition generated by the wave maker is programmed according to Eq.(\ref{surfaceelevation}). Each particular soliton solution requires a special choice of the carrier parameters.  The initial steepness of the carrier, defined as $\varepsilon_0=a_0k_0$, plays a key role in the experiment, with higher steepness $\varepsilon_0$ yielding more rapidly evolving dynamics.  On the other hand, wave breaking defines a threshold steepness value beyond which the excited soliton solution will break before reaching its maximal amplitude.  Our experiments indicate that breaking of the two-soliton and the three-soliton packets starts at initial steepnesses of $0.10$ and $0.05$, respectively and we thus kept the steepnesses below these values. The initial conditions in the experiments were of course chosen to ensure that the evolution dynamics of interest and the maximal amplification of the higher-order soliton were captured within the tank dimensions.  To observe maximal amplification within the tank, the  amplitude and steepness are respectively: 5 mm and 0.08 (for $N = 2$ soliton), and 2 mm and 0.04 (for $N = 3$ soliton).  Note that these values are far from the wave-breaking limit. 

The experimental results showing the measured evolution of these multi-solitons are shown in Fig. \ref{evo}. The upper panel shows the evolution of the two-soliton solution while the lower panel shows the three-soliton solution. As expected from NLS dynamics for the chosen initial conditions, the wave packet undergoes initial compression until it reaches its maximum amplitude at 8 m (for $N=2$ case) and 9.6 m ($N=3$ case).  It is important to note that gauges are at 20 cm intervals in experiments, which determines resolution in the recorded evolution.  

% Figure 3
\begin{figure}[ht]
\centering
\includegraphics[width=6.8cm]{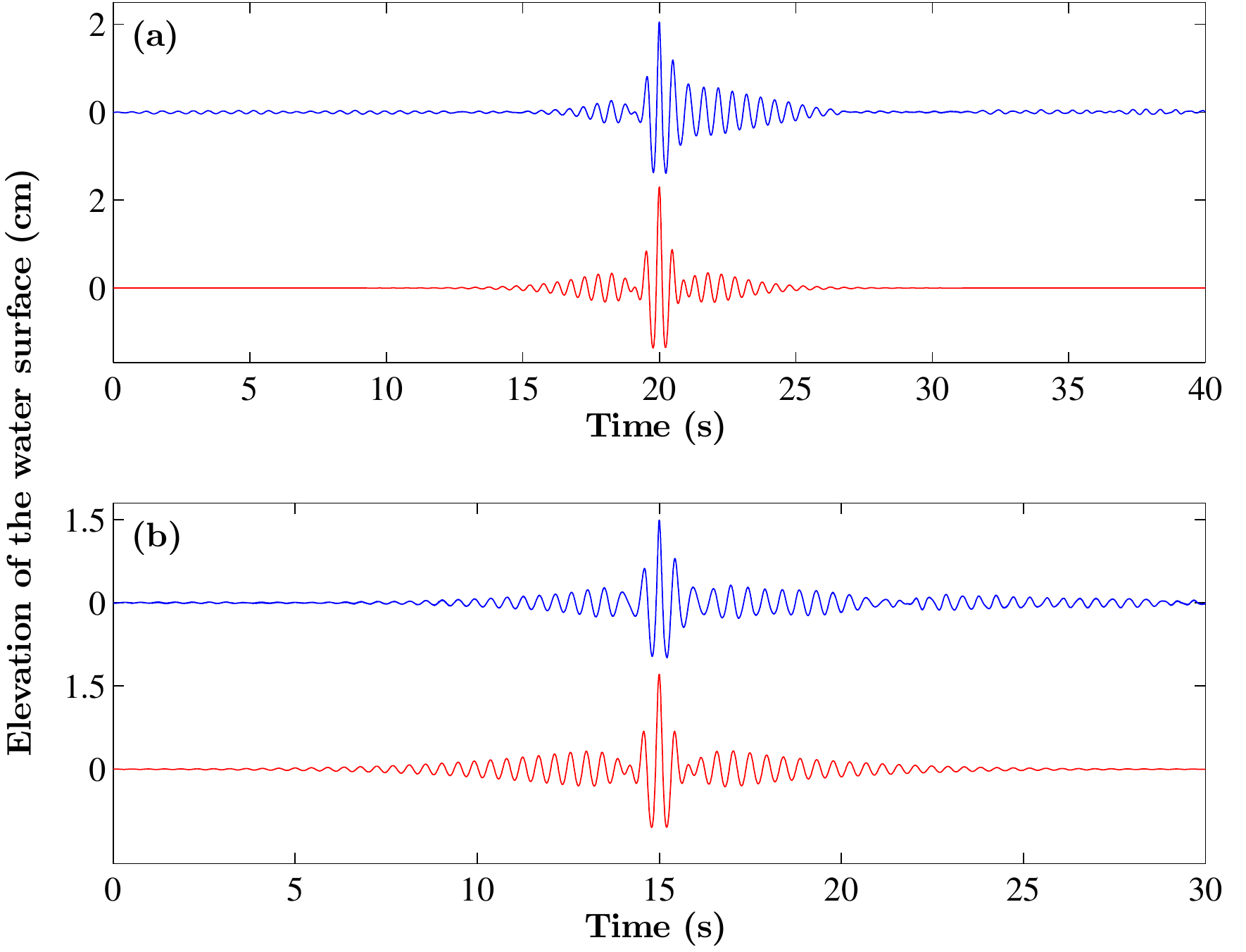}
\caption{(Color online) Comparison of the measured wave train (upper blue curves) with the corresponding NLS simulation (lower red curves) at the point of maximal measured wave amplification for: (a) the two-soliton breather with carrier parameters $a_0=5$ mm, $\varepsilon_0=0.08$ recorded at 8 m and (b) the  three-soliton breather with carrier parameters $a_0=2$ mm, $\varepsilon_0=0.04$ recorded at 9.6 m. The spatial deviations between the NLS simulation and the experiments are 1 and 3 cm for the $N=2$ and $N=3$ soliton, respectively.}\label{comparison}
\end{figure}

\begin{figure}[h]
\centering
\includegraphics[width=\columnwidth]{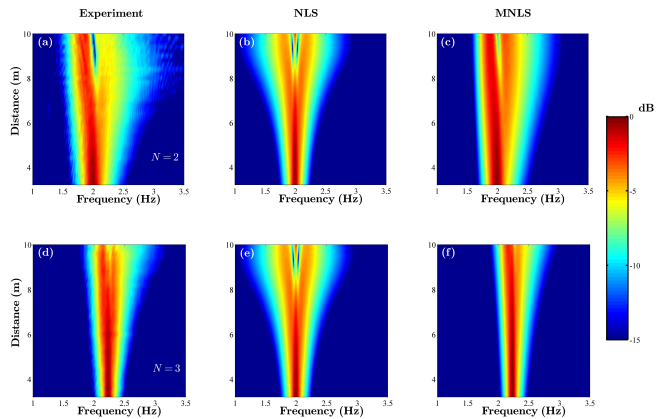}
\caption{(Color online) Spectral evolution of the $N=2$ (top) and $N=3$ (bottom) soliton. (a) \& (d) Experimental evolution calculated from the measured envelopes using the Hilbert-transform, (b) \& (e) simulated evolution using the NLS and (c) \& (f) simulated evolution using the MNLS.}\label{spectra_envelope}
\end{figure}

To quantitatively interpret these results in terms of the expected properties of higher-order solitons, we compare the experimentally observed wave groups at their maximal amplitude with that of the higher order soliton solutions simulated from the NLS. To this end, Fig. \ref{comparison} shows the measured profile (blue line) together with the corresponding simulated soliton solution (red line) at the spatial coordinate of maximal amplitude and excellent agreement is indeed observed for both the second- and third-order soliton. However, we can also see discrepancies in the form of a slight asymmetry in the experimental data. Such asymmetry is in fact expected from symmetry-breaking induced by higher-order nonlinear terms which become non-negligible when the wave group spectral bandwidth increases significantly during the phase of compression of higher-order solitons.  
Indeed, the propagation of deep-water waves (and optical pulses) can be modelled by the NLS only if the wave group or pulse envelope spectrum is narrow enough so that perturbations to the dispersive and nonlinear terms can be neglected.  In optics, such perturbations that can be included in the frame of a generalized nonlinear Schr\"odinger (GNLS) model have been shown to play a central role in the continuous spectral broadening referred to as  supercontinuum \cite{dudley2006supercontinuum}. For the case of deep-water waves, the physics of higher-order perturbations can be accounted for by using the MNLS, also sometimes referred to as the Dysthe equation, which includes additional terms compared to the NLS \cite{dysthe79}, and which can be considered as a hydrodynamic equivalent of the GNLS in optics. Of course the optical GNLS and the MNLS are not mathematically identical, but the models are physically analogous in that they both exhibit a term which represents the frequency-dependence of the nonlinearity as well as a delayed nonlinearity responsible for a frequency down-shift (representing the induced mean flow of the wave train in hydrodynamics \cite{DiasKharif}). Although the magnitude of these terms is different in both models, they induce similar physical perturbations to the narrowband NLS model.
The influence of higher-order perturbations on the spectral evolution of the $N=2,3$ solitons is confirmed in Fig. \ref{spectra_envelope} where we compare the experimental spectral evolution with the corresponding simulated evolution from both the NLS and MNLS.  The experimental spectra were calculated from the recorded wave envelopes using the Hilbert-transform. We see how the spectral evolution as predicted from the NLS is perfectly symmetrical around the carrier frequency and agrees well with the experiments until the stage of maximum compression where asymmetry is manifested. On the other hand, results obtained from the MNLS model do show asymmetry towards higher frequencies, in excellent agreement with the experimental observations. This result is important as it confirms the ability of the MNLS to describe accurately the propagation of 1D deep-water waves for steepness values below the wave-breaking threshold.
% Figure 4

\begin{figure}[h]
\centering
\includegraphics[width=\columnwidth]{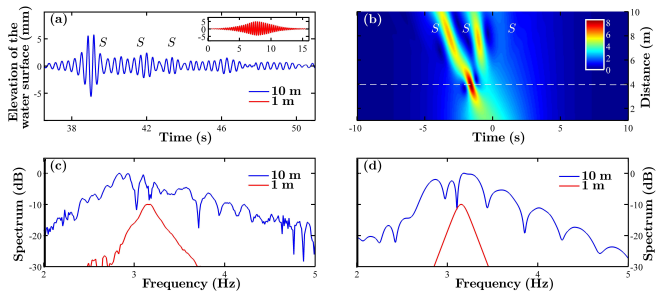}
\caption{(Color online) (a) Experimentally measured surface elevation at 1 m (inset) and 10 m from the flap with initial condition $4\operatorname{sech}(T)$ at $x = 0$. The experimental carrier parameters are $a_0=1$ mm  and $\varepsilon_0=0.04$.  The three largest fundamental solitons ejected from the fission are marked by the letter $S$. (b) MNLS simulation of the wave train envelope in the frame of reference moving at group velocity. The white dashed line indicates the theoretically calculated fission distance. (c) Experimental spectra at 1 m (red) and 10 m (blue).  (d) Corresponding results from MNLS simulations.}\label{fission}\end{figure}

In an optical context, the propagation of $N$-soliton solutions can lead to the generation of a broad supercontinuum when perturbation arising from higher-order dispersion and nonlinearity becomes important and split the initial bound state into fundamental individual solitons of different amplitudes and durations that separate with propagation, a mechanism generally referred to as soliton-fission \cite{Zakharov1984,kodama1987nonlinear,Clamond,dudley2006supercontinuum}. We next proceed to demonstrate that the very same phenomenon can also manifest itself in hydrodynamics. To this end, we increase the nonlinearity in the dynamical system in order to increase the spectral bandwidth of the wave group at the stage of maximum compression so that higher-order perturbations to the NLS becomes even more significant and break the initial bound-state. The symmetry-breaking caused by the higher-order terms then results in soliton fission with the individual solitons travelling at different velocities and eventually spreading across the wave group just as in the optical case \cite{dudley2006supercontinuum,GrossManassah,TranBiancalana}. In hydrodynamic propagation, increasing the nonlinearity can be achieved either by increasing the order of the launched multi-soliton or by increasing the carrier-steepness. We favoured the former in our experiment because increasing the steepness can cause the initial wave to break. Launching a $N=4$ soliton ($4\operatorname{sech}(T)$) into the tank with a carrier-amplitude $a_0=1$ mm and a carrier-steepness value of $\varepsilon_0=0.04$ we observe clear signatures of soliton fission into distinct fundamental solitons as seen in Fig. \ref{fission} (a), which shows the temporal amplitude of the wave group at the beginning (inset) and end of the tank. Correspondingly to the fission, we observe a quasi-continuous spectrum at the end of the wave tank as shown in Fig. \ref{fission} (c) similar to an optical supercontinuum. The hydrodynamic and optical analogy is further confirmed by noting that the distance at which fission occurs in the wave tank scales using the same criteria as that derived for pulses in optics. Specifically, the fission process is triggered at the distance of maximum temporal compression where the effect of higher-order perturbations are more pronounced and which can be approximated by $L_D/N$ where $L_D$ is the dispersive length. The dispersive length is simply given by the temporal width of the wave group at the input divided by the dispersion coefficient of the NLS, and in the water wave case is equal to $L_D=g/2T^2_0$ implying a fission length of about 4 m for our experiment. In principle, the fission could also be triggered by significant noise amplification but we have carefully checked through numerical simulations that the noise influence is negligible here and that the fission can be unambiguously attributed to higher-order perturbations to the NLS. The fact that we observe fission in the hydrodynamic case so apparently makes it clear that mathematical differences in the NLS perturbation terms relative to the optical system are not physically significant: it is because the higher-order perturbations break the symmetry of the integrable NLS that fission occurs, and this also inevitably leads to significant permanent spectral broadening with asymmetry. This is an important observation that illustrates how systems governed by the NLS are very sensitive to the presence of perturbations that are likely to break any bound-state into its fundamental constituents.  In order to highlight the crucial of the higher-order perturbations in the fission process we have performed numerical simulations using the extended MNLS, see Fig. \ref{fission}(b,d). We see excellent agreement with the output experimental spectrum. In particular fission at $L_D/N$ is clearly observed in the simulated temporal evolution (Fig. \ref{fission}(b)) and the magnitude of the spectral broadening is correctly predicted. 
% Figure 5

Of course there are some differences between the optical and hydrodynamic supercontinuum as we observe. Firstly, the difference in the linear dispersion properties between the optical and hydrodynamic systems means that we cannot satisfy a phase-matching condition for narrowband soliton-dispersive wave radiation in hydrodynamics as usually seen in optics \cite{dudley2006supercontinuum}.  Secondly, within the MNLS regime, the frequency down-shift experienced by solitons is weaker for water solitons than for optical solitons.  A direct consequence is that after the fission where solitons are ejected and separate with further propagation, the spectral broadening essentially ceases. This is in contrast with the optics case where the strong down-shift of the individual solitons can extend significantly the supercontinuum bandwidth towards the lower frequencies.

In conclusion, we have reported the observation of multi-soliton breathers on zero-background in hydrodynamics. The measured maximal wave amplitudes are in very good agreement with the analytical solutions of the NLS, and discrepancies are due to the higher-order effects which can be accounted for by the MNLS.  When the nonlinearity of the system is increased, higher-order perturbations break the multi-soliton bound state into fundamental solitons. Such a soliton fission mechanism is associated with extended spectral broadening along the flume leading to the generation of a water wave supercontinuum similar to that observed in optics.

These results not only reveal yet another correspondence between the dynamics of 1D wave tanks and fiber-optic systems, but most importantly they confirm that soliton fission and supercontinuum generation are likely to be universal phenomena encountered in a wide range of nonlinear systems governed by perturbed NLS-type equations. Naturally, it is also important to stress that, in the absence of such higher-order perturbations, fission will not occur as the system is perfectly integrable in this case. We anticipate that these results will motivate not only further studies in hydrodynamics to observe nonlinear interactions of surface gravity waves, but also studies in other nonlinear systems governed by similar type of NLS equations.

N.H. and N.A. acknowledge the support of the Volkswagen Stiftung. M.O. was supported by EU, project EXTREME SEAS (SCP8-GA-2009-234175) and ONR grant N000141010991. M.O. acknowledges Dr. B. Giulinico for interesting discussions. G.G. acknowledges the Academy of Finland (grants 130099, and 132279). J.M.D. is supported by the ERC project MULTIWAVE. N.A. acknowledges partial support of the Australian Research Council (Discovery Project No. DP110102068). N.A. is a winner of the Alexander von Humboldt Award.
\bibliographystyle{apsrev}
%\bibliography{references}

\end{document}